\documentclass[5p]{elsarticle}
\usepackage{graphicx}
\usepackage{booktabs}
\usepackage{hyperref}

\journal{Computer Physics Communications}
\biboptions{sort&compress}

\makeatletter
\def\ps@pprintTitle{%
 \let\@oddhead\@empty
 \let\@evenhead\@empty
 \def\@oddfoot{\footnotesize\copyright\ 2017, Licensed under the Creative Commons  CC-BY-NC-ND 4.0 license, \url{http://creativecommons.org/licenses/by-nc-nd/4.0/}\hfill}%
 \let\@evenfoot\@oddfoot}
\makeatother

\begin{document}
\begin{frontmatter}
\title{A hybrid algorithm for parallel molecular dynamics simulations} 
\author[lumath,mirarco]{Chris M. Mangiardi}
\ead{cmangiardi@mirarco.org}

\author[lumath,luphysics]{R. Meyer\corref{cor1}}
\ead{rmeyer@cs.laurentian.ca}

\cortext[cor1]{Corresponding author}

\address[lumath]{Department of Mathematics and Computer Science, Laurentian University, 935 Ramsey Lake Road, Sudbury, ON\ \ P3E 2C6, Canada}
\address[mirarco]{MIRARCO Mining Innovation, Laurentian University, 935 Ramsey Lake Road, Sudbury, ON\ \ P3E 2C6, Canada}
\address[luphysics]{Department of Physics, Laurentian University, 935 Ramsey Lake Road, Sudbury, ON\ \ P3E 2C6, Canada}

\begin{abstract}
This article describes algorithms for the hybrid parallelization and SIMD vectorization of molecular dynamics simulations with short-range forces. The parallelization method combines domain decomposition with a thread-based parallelization approach. The goal of the work is to enable efficient simulations of very large (tens of millions of atoms) and inhomogeneous systems on many-core processors with hundreds or thousands of cores and SIMD units with large vector sizes. In order to test the efficiency of the method, simulations of a variety of configurations with up to 74 million atoms have been performed. Results are shown that were obtained on multi-core systems with Sandy Bridge and  Haswell processors as well as systems with Xeon Phi many-core processors.
\end{abstract}


\end{frontmatter}
\section{Introduction}
Molecular dynamics is an important tool for the computational modelling of materials. On high-performance computers, large-scale molecular dynamics simulations are routinely run on large numbers  of processors in parallel, using significant amounts of CPU time. In order to allow simulations of large systems and to minimize the waste of valuable resources, it is therefore important that molecular dynamics programs are able to utilize the full computational power of modern CPUs. 

Recent increases of CPU power were achieved mainly through a larger number of processing cores per chip and increases of the vector size of SIMD (single instruction, multiple data) units. SIMD units allow a processor to perform the same operation on all elements of a short vector in parallel. The next step in this evolution will be many-core CPUs with several dozens, hundreds or thousands of processing cores \cite{Vajda:11}. The parallelism of these processors will be further enhanced through large SIMD vectors. (Current many-core processors such as Intel's Xeon Phi processors use 512 bit SIMD vectors \cite{Jeffers:13,Jeffers:16}. It is unclear whether this size will increase in the near future). Programs will only be able to unlock the full power of coming generations of processors if they use highly scalable parallel algorithms that run efficiently on a large number of processors in parallel and if they are able to take advantage of SIMD units. In addition to this, future CPUs might require that programs be able to adjust to  changes in the numbers of processing cores due to thermal management or graceful degradation. 

For molecular dynamics simulations on distributed parallel systems, there are three widely used methods for the distribution of the workload over the available processors: particle decomposition, force decomposition and domain decomposition. These methods are well described in Ref.~\cite{Plimpton:1995fc}. For large-scale simulations with short-ranged potentials, domain decomposition is the best of these methods since it minimizes the amount of data which needs to be communicated between processors.

The Achilles' heel of domain decomposition in molecular dynamics simulations is the workload balance. Domain decomposition divides the volume of the simulated system into subvolumes and assigns each subvolume to one processor. If the computational work required by the subvolumes differs substantially, the workload is not balanced and the simulation becomes inefficient. This happens in systems with considerable density fluctuations (e.g. porous systems) or multicomponent systems where the computational work required by the components differs.

On shared-memory systems, molecular dynamics simulations can also be parallelized using thread-based parallel APIs like OpenMP or Threading Building Blocks. If this route is followed, care must be taken to avoid the occurrence of race conditions when two processors update the data of the same particle simultaneously. This is particularly important if Newton's third law is exploited so that the force between two particles is evaluated only once. Synchronization measures to avoid these race conditions can significantly reduce the parallel efficiency of a molecular dynamics program. 

In this work, we describe the implementation of a hybrid algorithm for molecular dynamics simulations which employs a two-level parallelization method. The first level uses domain decomposition to subdivide the system into smaller subvolumes. The second parallelization level parallelizes the computations within the subvolumes with the help of a recently proposed task-based parallel algorithm for molecular dynamics simulations on shared-memory systems \cite{Meyer:2013im}. This algorithm --- named the \textit{cell-task method} --- has been designed to overcome the problems that domain decomposition faces when it is applied to inhomogeneous systems. The hybrid algorithm thus extends the range of the cell-task method from single shared memory systems to distributed systems with multiple shared-memory compute nodes. 

In addition to the hybrid parallel algorithm, we also describe and discuss our approach to the SIMD vectorization of the force calculation and neighbor-list generation in molecular dynamics simulations. The SIMD vectorization of molecular dynamics simulations with short-range forces is made difficult by the irregular memory access pattern of such simulations. Our approach to this problem is a blocked algorithm which is different from the clustering algorithm used in earlier work~\cite{Friedrichs:2009ir,Pall:2013gb}.

We tested both, the hybrid algorithm and the vectorization algorithm, on typical multi-core systems with Xeon processors as well as systems with Xeon Phi processors. Xeon Phi is the brand name for a series of many-core processors developed by Intel which feature a large number of cores per processor and 512 bit SIMD units \cite{Jeffers:13,Jeffers:16}. 

Many parallelized molecular dynamics programs like LAMMPS \cite{Plimpton:1995fc}, NAMD \cite{Phillips:2005ia}, GROMACS \cite{Berendsen:1995dd}, DL\_POLY \cite{Todorov:2006ee}, or MOLDY \cite{Ackland:2011kw} are freely available and widely used. The purpose of our work is not to add another program to this list.
Our principal objective is to lay the ground work for future revisions of such general purpose molecular dynamics simulation programs. As computer architectures evolve, the pressure will rise to adapt molecular dynamics programs so that they make the best use of the available computing resources. We hope that our work will provide useful information in this direction.  Some of our results presented here have been discussed earlier in Refs.~\cite{Meyer:2015ky,Mangiardi:2016a,Mangiardi:2016b}.

\section{Algorithms}
%
%
\subsection{Hybrid parallelization}
The hybrid algorithm described in this article combines the message-passing based domain decomposition method with the thread-based cell-task method. In standard implementations of the domain decomposition method a single processor core running a single thread performs the calculations related to the atoms in one subdomain. The hybrid algorithm, instead, employs for each sub-domain a team of threads running on multiple cores.

We have implemented the hybrid algorithm within our existing molecular dynamics program MDNTP which was already used to build the cell-task method. MDNTP is a general purpose parallel molecular dynamics code for large-scale simulations with short-range forces. It has been used successfully in a number of studies (e.g. Ref.~\cite{Meyer:1998iw,Derlet:2001is,Meyer:2003gt,Meyer:2011ha,Meyer:16a}). Compared to the widely used molecular dynamics programs mentioned in the preceding section, MDNTP is a relatively small code which facilitates the incorporation and testing of new features.

The original version of MDNTP included domain decomposition as its parallelization method. Recently,  the cell-task method~\cite{Meyer:2013im} was added to the program as an alternative parallelization method for simulations on a single (shared-memory) node. The hybrid algorithm combines the two methods so that both parallelization schemes can be used simultaneously.

The cell-task algorithm divides the force calculations and the construction of neighbor lists into a large number of small tasks. To do this, the simulation volume is divided into small cells and a task consists of the calculation of the force contributions or the generation of the neighbor lists for all atoms in one cell. The tasks are then executed by a pool of worker threads according to a conditional schedule. The schedule is built in such a way that simultaneously running tasks will never access the same particle. This way, the cell-task method avoids the need for locks or other synchronization mechanisms. Ackland \textit{et al.} have described a similar method using locks~\cite{Ackland:2011kw}. 

By its design, the cell-task method is compatible with domain decomposition. For the algorithm it does not matter whether the volume comprises the complete simulation cell or only a part of it. Due to the compatibility between the two methods, only three areas of the code had to be changed in order to make the hybrid algorithm work. First, the implementation of domain decomposition in MDNTP splits the calculations of forces and the construction of neighbor lists into an inner and an outer part. The inner part involves only the atoms in the domain of the processor (or node) while the outer part deals with the interactions between particles in different domains. This makes it possible to communicate particle data asynchronously at the same time as the calculations of intra-domain interactions. For example, the calculation of the forces from a pair potential follows the general algorithm below:

\begin{enumerate}
\item Assemble coordinates of particles near domain boundaries in send buffers.
\item Initiate sending of the outgoing buffers.
\item Initiate reception of incoming coordinates.
\item Compute force contributions from intra-domain interactions.
\item Wait for completion of the receive operations.
\item Compute force contributions from inter-domain interactions.
\item Wait for completion of the send operations. 
\end{enumerate}

Except for the explicit wait operations in steps 5 and 7, no synchronization is required between the domains. In particular, there is no barrier between the steps. In our implementation, the steps of this algorithm are executed in sequence for each domain by the responsible processors making use of parallelism within each step. Alternatively, it would be possible to implement the first three steps as tasks and execute them in parallel with step 4. This might, however, increase the wait time in step 5 since the send operation will start later. It would also be possible to execute step 3 before steps 1 and 2 which might increase the performance on some systems. 

In order to make the algorithm described above compatible with the cell task method, very few changes were required. First, a second set of tasks was implemented for the calculation of the intra-domain forces in step 6 and the construction of the corresponding neighbor lists. Second, the assembly of the message buffers in step 1 was parallelized to take advantage of the availability of multiple processors for each domain. Although the single-threaded code from the original implementation of this step would still work and only a small amount of time is spent in this step, the assembly of the buffers might become a bottleneck if many processor cores were used.

The parallelization of the message buffer assembly is not overly complicated. Conceptually, the code consists of a double loop over the different buffers --- one for each neighboring domain --- and the particles whose data go into each of these buffers. A parallelization of the outer loop over the buffers is excluded here since the number of available processors can be vastly larger than the number of buffers. A parallelization of the inner loop is more reasonable, but it means a repetition of the overhead that comes with the parallelization of a loop.  To avoid unnecessary overhead, we treated the set of individual buffers as one large continuous buffer, effectively collapsing the double loop into one. This way, only one loop needs to be parallelized each time the buffers are assembled. 

Finally, the code responsible for the identification of particles that have moved out of a domain and the particles that are near to a domain boundary was updated. As with the code for the assembly of the message buffers, the single-threaded code from the original implementation of the domain decomposition method would still be usable at the price of reduced parallelism. This code consists mainly of a single loop that runs over all particles, eliminates particles that have moved across a domain boundary from the main particle set  and groups the remaining particles into 27 areas. The areas are later used to determine which particles belong to which message buffer. The problem with the parallelization of this loop is that all processors are constantly changing the contents of the data structures of the 27 areas. This requires synchronization of the access to the data structures and leads to massive problems with so-called \textit{false sharing} \cite{HPC}. To avoid the synchronization as well as \textit{false sharing}, we introduced thread-local versions of the 27 data structures so that during the loop each thread uses its own set of data structures. After the loop has finished, the contents of the thread-local data structures are then merged. Although this merging is done by a single thread, this should have a negligible effect on the efficiency as the time for this operation does not increase with the number of particles.   

With the three code changes described in the preceding paragraphs the most time consuming parts of the hybrid algorithm are multithreaded. Of the remaining code, some sections might also benefit from multithreading. Many of these sections are, however, either not parallelizable or their parallelization would require significant changes to central data structures.  Since the amount of time spent in these sections of the code is very small, an impact on the results presented in this work can be excluded.    

%
%
\subsection{SIMD vectorization}
The problem of molecular dynamics simulations with short-range forces is not well suited for SIMD vectorization. The reason for this can be seen from the typical code for the force calculation for a pair potential $\Phi$:
 
\begin{enumerate}
\item \textbf{for} $i$ in all $particles$ \textbf{do}:
\item \hspace{2em}\textbf{for} $j$ in $neighbors(i)$ \textbf{do}:
\item \hspace{4em}$\mathbf{R} \leftarrow \mathbf{r}_i - \mathbf{r}_j$
\item \hspace{4em}$d \leftarrow |\mathbf{R}|$
\item \hspace{4em}$\mathbf{F} \leftarrow -\frac{1}{d}\frac{\mathrm{d} \Phi}{\mathrm{d} r} \mathbf{R}$
\item \hspace{4em}$\mathbf{f}_i \leftarrow \mathbf{f}_i + \mathbf{F}$
\item \hspace{4em}$\mathbf{f}_j \leftarrow \mathbf{f}_j - \mathbf{F}$
\item \hspace{2em}\textbf{end for}
\item \textbf{end for}
\end{enumerate}

Since the neighbors $j$ of particle $i$ are not stored consecutively in memory, the components of the position $\mathbf{r}_j$ of particle $j$ and of the force $\mathbf{f}_j$ acting on it are also not stored in consecutive memory locations. For this reason, the data for multiple iterations of the inner loop cannot be loaded or stored through normal SIMD instructions that access consecutive memory locations. The Xeon Phi processor instruction sets contain gather and scatter instructions which load from or store to scattered locations. However, although these instructions reduce some of the overhead, they do not change the fact that multiple memory transactions are required for these instructions. Effectively, gather and scatter instructions still imply serialized memory accesses.  

The fact that the loading of $\mathbf{r}_j$ at the beginning of the inner loop and the storing of $\mathbf{f}_j$ at its end cannot be SIMD vectorized does not preclude the vectorization of the rest of this loop. Once the components of $\mathbf{r}_j$ for successive loop iterations have been loaded into SIMD registers (through a gather instruction or other means), the calculations can proceed using SIMD instructions. In our SIMD vectorization we still chose a slightly different path.

The nested loops in the force calculation shown above can be interpreted as one loop running over all pairs of neighboring atoms. In our work, we implemented the loop in this way and then used a blocked algorithm to execute smaller chunks of the loop. With this approach the force calculation for the pair potential $\Phi$ looks like this:

\begin{enumerate}
\item \textbf{while} not all pairs processed \textbf{do}:
\item \hspace{2em}fillBuffers($bx$, $by$, $bz$)  
\item \hspace{2em}calcDistances($bx$, $by$, $bz$, $bd$)
\item \hspace{2em}calcDerivatives($bd$, $bf$)
\item \hspace{2em}storeBuffers($bf$, $bx$, $by$, $bz$)
\item \textbf{end while}
\end{enumerate}

In this scheme, $bx$, $by$, $bz$, $bd$, $bf$ are properly aligned short arrays acting as buffers for one block of particle pairs. The function fillBuffers() fills three buffers with the distance components of the next set of particle pairs. The function stops when the buffers are full or there are no more particle pairs left. The function calcDistances() then fills $bd$ with the particle distances and calcDerivatives() stores the values of $-\frac{1}{d}\frac{\mathrm{d}\Phi}{\mathrm{d} r}$ for the pairs in the buffer $bf$. Finally, storeBuffers() calculates the inter-particle forces and updates the particle data $\mathbf{f}_i$ and $\mathbf{f}_j$. Only the functions fillBuffers() and storeBuffers() perform scattered memory accesses. The other functions only make consecutive accesses to the short buffers.

The blocked algorithm is not limited to pair potentials. We only use the example of pair potentials to describe the blocked algorithm due to their simplicity. The scheme can easily be extended to more complicated interactions such as the many-body interactions used for the benchmark simulations described in Sec.~\ref{SecDetail} and~\ref{SecRes}. For many-body interactions the blocked algorithm is used independently for the calculation of the localization functions and the calculation of the forces.

There are multiple reasons why we chose a blocked algorithm instead of a direct vectorization of the loop over the neighbors of each atom. First, our blocked algorithm separates vectorizable and non-vectorizable operations. In addition to this, the shorter loops reduce register pressure. If all calculations are done in one loop, it may not be possible to keep all required parameters in registers so that additional memory transfers are necessary during each loop iteration. Since the blocked algorithm results in a series of leaner loops, it is often possible to keep all parameters in registers throughout one loop. Parameter reloads then occur only at a reduced frequency  between the shorter loops. Altogether, the breakup of the calculations into a series of short loops improves the quality of the generated machine code since it makes it easier for the compiler to analyze the loop bodies for vector dependencies and to identify vectorizable loops.  

In addition to an improvement of the code quality, the blocked scheme makes it possible to add non-vectorizable operations to the force calculation. For example, the update of a radial distribution histogram during the force calculation  cannot be vectorized, which might stop the compiler from vectorizing the loop at all. In the blocked scheme, a call to a function updating the histogram for all pairs currently in the buffers can be added to the outer loop over the blocks. The fact that the histogram evaluation is not vectorized does not affect the vectorization of the other loops. Similarly the evaluation of optional computations in the inner force loop can be made more efficient in the blocked algorithm. To do this, the evaluation of such an optional feature is moved into its own small loop and the loop is only invoked if required. This avoids a conditional statement in the innermost loop.  

Another advantage of our blocked algorithm is that it reduces the number of cases where the full width of the vector registers cannot be used. Since the blocked algorithm collapses the double loop over particles and their neighbors into a single loop, only the last iteration of the last block might use a partial vector register. Without the collapsed loops, partial register filling may occur once for every particle in the system.

The SIMD vectorization of the force calculations becomes more complicated if the simulation contains multiple atom types (elements). In this case, the calculations of the derivative of the potential require parameters which depend on the types of the interacting particles. Since the parameters are not stored consecutively, this entails more scattered memory accesses which reduce the efficiency of the vectorization. In order to limit the impact of multiple atom types, we implemented the force calculation in such a way that parameter reloads are only used when multiple elements are used.   

The Xeon Phi's architecture supports masked vector instructions. We use this to reduce the negative impact of subsequent reloads of parameters. Instead of reloading parameters indiscriminately for each calculation, the interaction type of the parameters currently loaded is kept in a variable and reloads are only carried out if the parameter changes. The compiler can translate this into masked operations which reduce the number of elements in the vector registers which need to be reloaded. As will be shown below, this trick improves the vector performance for systems where changes of the interaction types occur only in a small part of the system.   

It has been argued that the SIMD vectorization of molecular dynamics simulations requires extensive rewriting of the code and that this effort might have to be repeated for future hardware generations \cite{Needham:2016hc}. In our implementation, however, only two sections of the code were affected by the vectorization and the work required to vectorize different types of interaction potentials was kept low through the use of C++ templates. Further, our blocked algorithm is not system specific and works on a high level by rearranging the loop structure of the code. All system or architecture specific details are handled by the compiler. The only system dependent parameter is the length of the buffers which can be set at runtime. 

%
%
\section{Details of the benchmark simulations}\label{SecDetail}
\subsection{Test configurations}\label{SecCfg}
The parallel speedup of a molecular dynamics simulation depends not only on the program used for the calculations but also on the properties of the simulated system. The potential describing the interatomic forces, the number of particles in the system, homogeneity of the system, presence of multiple elements and their distribution are some of the factors that can influence the speedup achieved in a simulation. For this reason, we determined the parallel speedup for a number of different benchmark configurations. Comparison of the results makes it possible to judge the impact of different factors and to identify the main limitations of a given architecture. In this section we describe the different benchmark configurations used in this work.

The type of potential from which the inter-atomic forces are calculated clearly has a strong influence on the speedups that can be obtained. In particular the success of the SIMD vectorization depends critically on the type of functions that need to be evaluated. The potentials used in this work are the tight-binding second-moment potentials for copper and gold by Cleri and Rosato~\cite{Cleri:1993dn} and the embedded-atom method~\cite{Daw:1984ix} potential for iron by Mendelev \textit{et al.}~\cite{Mendelev:2003co}. Both types of potentials have a similar functional form that can be written as
\begin{equation}
   U = \sum_{i} F_i(\rho_i)\;+\;\frac{1}{2}\,\sum_{i \ne j} \Phi_{ij}(r_{ij})
\end{equation}
where $U$ is the potential energy of the system, $F_i$ is the embedding energy for atom $i$, $\Phi_{ij}$ is a pair interaction between atoms $i$ and $j$, $r_{ij}$ is the distance between these atoms and 
\begin{equation}
	\rho_i = \sum_{j \ne i} \psi_{ij}(r_{ij})   
\end{equation}
is the localization function for atom $i$ which is calculated as a sum over the density functions $\psi_{ij}(r_{ij})$. 

For both potentials, the forces can be computed in three steps. In the first and second step the localization function $\rho_i$ and  the embedding energies $F_i$ are calculated for each atom, respectively. The third step evaluates the pair interaction $\Phi_{ij}$ and finishes the calculation of the forces. The first step and the third step require summations over all atoms and their neighbors. These steps can therefore be implemented by a pair algorithm as described above. The second step only requires a loop over all atoms and is usually very fast.

The difference between the potentials is due to the type of functions involved. The Cleri and Rosato potentials employ simple exponential functions for $\psi_{ij}$ and $\Phi_{ij}$ and a square root function for $F_i$. In contrast to this, the Mendelev potential uses piecewise definitions for the functions $\psi_{ij}$ and $\Phi_{ij}$ which hamper their vectorization.

\begin{figure}%
\centerline{\includegraphics[width=8.5cm]{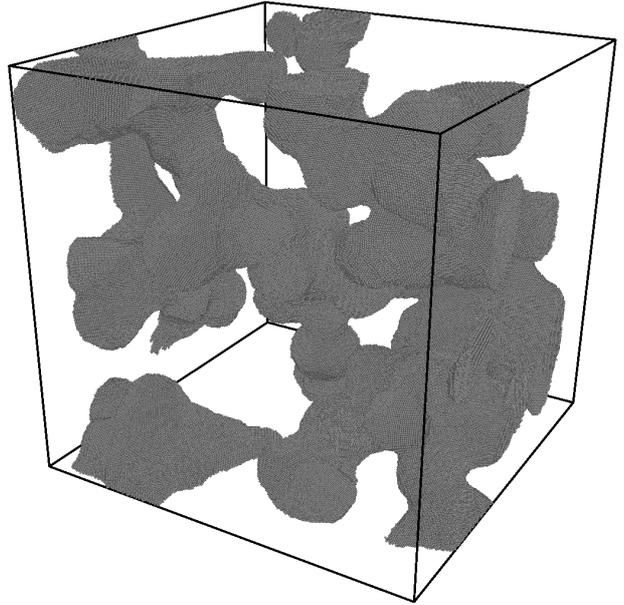}}
\caption{Test configuration \textit{Porous} representing porous copper. Reproduced with permission from Ref.~\cite{Meyer:2013im}.}%
\label{FigCubo}
\end{figure}

To study the behavior of the SIMD vectorization and the hybrid algorithm on a single node, we used a set of five different test configurations. \textit{Cu105} is a homogeneous bulk system consisting of $105\times 105\times 105$ cubic fcc cells containing 4,630,500 copper atoms. The system uses the computationally relatively simple copper potential by Cleri and Rosatto~\cite{Cleri:1993dn}. The same potential is used by a 1,992,220 atom configuration named \textit{Porous}. This configuration represents an inhomogeneous system of porous copper obtained from the sintering of copper nanoparticles (see Fig.~\ref{FigCubo}). The inhomogeneous density distribution of \textit{Porous} makes this system challenging for the domain decomposition method of parallelization. 

\textit{Fe126} is another homogeneous, cubic  bulk system containing 4,000,752 iron atoms on a  $126\times 126 \times 126$ cell  bcc lattice. In contrast to the previous system, \textit{Fe126} uses Mendelev's potential for iron~\cite{Mendelev:2003co}. 

In order to test the influence of the presence of multiple elements on the vectorization, we built two multicomponent systems using the  Cu-Au potential from Ref.~\cite{Cleri:1993dn}. The main difference between the two systems is the spatial distribution of the elements. \textit{Alloy} is a cubic block of a disordered $\mathrm{Cu}_{75}\mathrm{Au}_{25}$ alloy with a total of 4,630,500 atoms on a $105\times 105\times 105$ cell fcc lattice. In contrast to this, \textit{Layer} consists of one layer of copper and one layer of gold with a total of  3,919,212 atoms. While \textit{Alloy} requires calculations of  interactions between all combinations of elements (Cu-Cu, Cu-Au and Au-Au) everywhere in the system, most of the volume of \textit{Layer} consists of a pure element. Thus, changes of the potential parameters only occur close to the interfaces between the two layers.

\begin{figure}%
\centerline{\includegraphics[width=8.5cm]{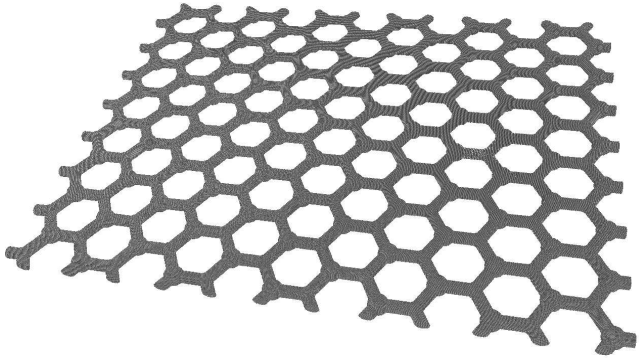}}
\caption{Phononic crystal configuration \textit{PnC} with 12.1 million copper atoms.}%
\label{FigHC}
\end{figure}
In order to see the performance of the hybrid algorithm over multiple nodes, we used two types of configurations. The first type consists of homogeneous copper bulk systems. To study the strong scaling behavior, a configuration \textit{Cu144} with $144\times 144\times 144$ fcc cells (11.9 million atoms) was used. For the weak scaling experiments, systems of different sizes were generated so that each node simulated 4.6 million atoms  ($105\times 105\times 105$ cells). In addition to the bulk systems, the multi-node benchmark simulations employed a set of phononic crystal (PnC) systems made from copper nanowires and nanoparticles. Figure~\ref{FigHC} shows as an example the 12.9 million atom configuration \textit{PnC} used to measure the strong scaling. The weak scaling test used similar configurations so that each node covered a domain with 3.8 million atoms. For the phononic crystal systems, the boundaries of the domains only cut some of the nanowires. Therefore, the number of atoms to be communicated between nodes is much lower in the case of the phononic crystal systems than in the case of the copper bulk systems.

The phononic crystal systems were inspired by similar simulations of silicon phononic crystals~\cite{Meyer:16a}. Although it was not yet available for the work in Ref.~\cite{Meyer:16a}, the hybrid algorithm was implemented to enable simulations of this kind of systems. It would be inefficient to simulate the configurations with domain decomposition alone since for  a large number of processors there is no way to obtain a similar number of atoms in each domain.  
 
 %
 %
\subsection{Test systems}
The benchmark simulations in this work were carried out on three different types of computers. The first machine consisted of a number of nodes with two 2.7 GHz Intel Xeon E5-2680 CPUs. This type of CPU is based on the Sandy Bridge microarchitecture and has 256 bit (4 double precision values) vector registers supporting Intel's AVX instruction set. Each node had a total of 16 CPU cores.

The second type of computers had dual 2.6 GHz Intel Xeon E5-2690 v3 CPUs based on the newer Haswell microarchitecture. These processors also have 256 bit vector registers but they support the AVX-2 instruction set. Each of the CPUs has 12 cores giving the nodes 24 cores each.  

In order to test our algorithm on many-core processors with a large number of cores, we also ran simulations on a machine containing a series of host nodes with attached Xeon-Phi coprocessors. Each host node  features dual Xeon E5-2670 CPUs (Sandy Bridge microarchitecture, 2.6 GHz) as well as two Xeon Phi 5110P coprocessors. These first generation Xeon Phi processors have 240 logical cores (60 physical cores supporting 4 hardware threads each) and 512 bit (8 double precision values) SIMD vector units. 

First generation Xeon Phi processors, known as Knights Corner (KNC), are only available as coprocessors that are connected to a host system through a PCIe bus. They can be used in two modes. In offloading mode, the main program runs on the host machine and delegates (offloads) parts of the calculations to the coprocessor. This model is very similar to typical GPU usage models. In native mode, a program can be run directly on a Xeon Phi coprocessor without participation of the host. All simulations described in this article used the native mode.

Intel recently released a series of second generation Xeon Phi processors called Knights Landing (KNL). At the end of our work we obtained access to a small development system with Xeon Phi 7210 processors. This processor type has 64 physical cores supporting up to 256 logical cores and it is only available as a stand-alone processor, not as a coprocessor. We performed some preliminary tests on this machine to complement our benchmarks. While this test system consisted of four nodes, there were differences in the configurations so that these were effectively two pairs of identical systems. All of our calculations were done on two nodes configured with Cluster Configuration set to ``Quadrant" and MCDRAM in ``Cache" mode. 

The implementation of hardware threads on Knights Corner coprocessors does not utilize Intel's hyper-threading technology (see for example page 17 in Ref.~\cite{Jeffers:13}). For this reason, we avoid the term hyper-thread in this article and use instead the generic term hardware thread.

\subsection{Benchmark simulations}
All simulations were carried out with similar simulation parameters. The simulations were run at a temperature $T=300\,\mathrm{K}$ using the Velocity-Verlet algorithm~\cite{AllenTildesley} with a time step of 2\,fs. Periodic boundary conditions were applied in all directions with the exception of the phononic crystal systems which used open boundary conditions along the z-axis. The neighbor lists were rebuild every ten simulation steps. Instead of a fixed rebuild interval, we could have rebuilt the lists whenever at least one atom had moved more than half of the difference between the cutoff radius used for the neighbor lists and the cutoff radius of the potential. However, with this dynamic approach the number of neighbor-list rebuilds would be different for each test configuration which would make it more difficult to compare the speedups of the configurations. 

For simulations of large systems, the time to load the initial configuration and to store the final configuration can be substantial. For this reason, we measured the execution time of the simulations excluding these I/O times as well as other overhead at startup and shutdown. Moreover, the measurement of the simulation times started only at the second construction of the neighbor lists, i.e, at the 10th step. The reason for this is that during the first construction of the lists a large amount of memory is allocated. The first construction of the lists therefore takes considerably more time than subsequent list generations. To exclude this one-time effect which is unimportant for longer simulations, time measurement started only at the 10th step when the neighbor lists were generated for the second time. Most simulations were run over a period of 1010 simulations steps in total so that the measured execution times covered 1000 steps.  Only some of the single-threaded simulations were run over a shorter period in order to keep the run times within acceptable limits. This affects the non-vectorized single-threaded simulations of the systems \textit{Cu105}, \textit{Porous}, \textit{Fe126}, \textit{Alloy} and \textit{Layer} as well as the vectorized single-threaded runs of \textit{Cu144} and \textit{PnC}. These simulations were only run over 510 simulation steps in total, measuring the execution times over 500 steps. Since these simulations are still far longer than even the longest multithreaded counterparts, this should not affect the quality of the measurement.  

The block size of the SIMD algorithm was determined for each architecture through some experimentation. We found that in general the algorithm is not very sensitive to the exact value and provides very similar results over a broad range of block sizes. As a rule of thumb the block size should be chosen such that all buffer arrays together fit in the L1 cache. We chose block sizes of 384 on the Xeon based architectures, 128 for simulations on the Knights corner systems and 256 on Knights Landing. The smaller block sizes on the Knights Corner processors reflect the fact that in this architecture four logical processor cores share the same L1 cache. The same is true for the Knights Landing architecture. For this architecture, however, experiments gave better results with the larger block size although the buffer arrays do not fit into the L1 cache. We believe that the reason for this is better hardware prefetching on the newer generation of Xeon Phi processors.     

The execution times of molecular dynamics simulations show some variation even between identical runs. The main reasons for this are system processes or other external events such as I/O requests. There also is a random element in the execution of the task schedule of the cell-task method. To reduce these variations and to get an idea of the magnitude of the variations, we repeated every simulation five times and calculated the average and standard deviation for the five repetitions.  

On the Xeon based systems, we ran the multithreaded simulations with one thread per available core. Since hyperthreading was turned off on these machines, the number of logical cores was equal to the number of physical cores resulting in 16 available cores on the nodes with Sandy Bridge CPUs and 24 available cores on the nodes with Haswell CPUs. 

On the Xeon Phi systems, the number of threads with which optimal performance is achieved depends strongly on the application. For this reason, we repeated all multithreaded simulations on these systems over a range of thread numbers and used the number of threads that gave the shortest time averaged over five identical runs. 
 
For simulations that make use of domain decomposition, the subdomains form a regular grid of $n_x\times n_y\times n_z$ subdomains. The numbers $n_x$, $n_y$,$n_z$ were chosen so that  the most compact form of the subdomains was maintained. For example, in a simulation with 24 domains the arrangement $4\times 3 \times 2$ was preferred over $8\times 3\times 1$. For the phononic crystal configurations only two-dimensional grids $n_x\times n_y\times 1$ were considered. On the Xeon Phi, Knights Corner coprocessors, Intel's implementation of MPI version 5.0.1 was used in combination with the tmi API for low-level communication. On the Xeon based systems, OpenMPI, version 1.6.2 was used with a QDR Infiniband interconnect on the Sandy bridge nodes and an FDR Infiniband interconnect on the Haswell nodes.

The simulation program MDNTP can be compiled with or without support for domain decomposition. The version without support for domain decomposition uses slightly leaner data structures and is somewhat faster on single node systems. For the benchmark simulations described in the following section, the version without support for domain decomposition was only used in the SIMD vectorization tests. All other simulations used the version with support for domain decomposition.

For the compilation of the simulation program we used Intel's C/C++ compiler version 17.0.1 except for the simulations using MPI on Knights Corner processors. The executable for these simulations was compiled with version 15.0.0 of this compiler due to restrictions in the availability of system libraries. The compilations used the optimization flags ``-O3 -ansi-alias -unroll-aggressive" for all files. The vectorized parts of the code were compiled with the additional flag ``-qopt-assume-safe-padding". To disable the generation of vector code we employed the flags ``-no-vec -no-simd".  

%
%
\section{Results and discussion}\label{SecRes}
%
%
\subsection{SIMD vectorization}\label{SecVec}
\begin{table*}
\caption{\label{TabVec}Parallel speedups with and without multithreading. $S_\mathrm{v,st}$ and $S_\mathrm{v,mt}$ denote the speedups achieved by vectorization with single-threaded and multithreaded execution, respectively. $S_\mathrm{mt}$ is the speedup achieved through multithreading without vectorization and $S_\mathrm{tot}$ is the total speedup resulting from vectorization and multithreading. The ideal speedups for each architecture are given in the last row.
 }
\centerline{\begin{tabular}{@{}lllllcllllcllll@{}}\toprule
&\multicolumn{4}{c}{Sandy Bridge} & \phantom{a} & \multicolumn{4}{c}{Haswell} & \phantom{a} & \multicolumn{4}{c}{Knights Corner}\\
\cmidrule{2-5} \cmidrule{7-10} \cmidrule{12-15}
&$S_\mathrm{v,st}$& $S_\mathrm{v,mt}$ & $S_\mathrm{mt}$ & $S_\mathrm{tot}$& &$S_\mathrm{v,st}$ & $S_\mathrm{v,mt}$ & $S_\mathrm{mt}$ & $S_\mathrm{tot}$& &$S_\mathrm{v,st}$ & $S_\mathrm{v,mt}$ & $S_\mathrm{mt}$ & $S_\mathrm{tot}$\\
\midrule
\textit{Cu105}    & 2.16  & 1.90 & 13.03 & 24.77 & 	& 1.97 & 1.68 & 17.87 & 30.10  &	&  3.12 & 2.90 & 131.64 & 381.11 \\
\textit{Porous}   & 1.91  & 1.80 & 12.99 & 23.42 &  	& 1.83 & 1.64 & 18.17 & 29.78  & 	&  3.05 & 2.74 &  \phantom{0}97.12 & 266.31 \\
\textit{Fe126}    & 1.60  & 1.42 & 11.82 & 16.77 &  	& 1.54 & 1.34 & 15.77 & 21.14  &  	& 1.82 & 1.73 & 129.01 & 223.47 \\
\textit{Alloy}      & 1.72  & 1.63 & 13.23 & 21.56 &   	&1.76 & 1.66 & 18.90 & 31.33  &  	&  2.92 & 2.53 & 122.09 & 309.46  \\
\textit{Layer}     & 1.98  & 1.84 & 13.10 & 24.11 &	& 1.79 & 1.67 & 18.46 & 30.88  &  	&  3.00 & 2.66 & 118.56 & 315.33 \\
\midrule
Ideal	 		& 4.00  & 4.00 & 16.00 & 64.00 &	& 4.00 & 4.00 & 24.00 & 96.00 &   	&  8.00 & 8.00 &  120.00 & 960.00 \\ 
\bottomrule
\end{tabular}}
\end{table*}
In Table~\ref{TabVec} we present the speedups obtained through SIMD vectorization and multithreading in our benchmark simulations on the three test systems. All speedups were calculated with respect to the execution time of a non-vectorized single-threaded simulation. The executables for non-vectorized simulations used exactly the same program code as the vectorized simulations. The vectorization was disabled during the compilation of the executables with the options ``-no-vec -no-simd" which prohibit the generation of SIMD instructions. Single-threaded simulations did not require separate executables. When the number of threads is set to one, the cell-task method falls back to a sequential algorithm that avoids the overhead of task scheduling. The ideal speedups given in Table~\ref{TabVec} reflect the degree of parallelism provided through the use of multiple CPU cores and/or SIMD vectorization.  For Knights Corner we assume twice the number of physical cores since on this architecture a single thread can issue instructions only every other clock cycle.

The observed speedups under single-threaded execution in Table~\ref{TabVec} range from 1.54 to 2.16 for the two Xeon architectures and from 1.82 to 3.12 for the Knights Corner processors. Theses values are considerably lower than the size of the vector registers which are 4 in case of the two Xeon architectures and 8 in case of the Xeon Phi. This reflects the fact that molecular dynamics simulations cannot be vectorized completely. Comparison of the speedups $S_\mathrm{v,mt}$ and $S_\mathrm{v,st}$  shows that multithreading slightly reduces the vectorization speedups. This is most likely due to the increased pressure on the memory system when all CPU cores are in use. The speedups $S_\mathrm{v,mt}$ obtained with multithreading nevertheless follow the same trends as their single-thread counterparts $S_\mathrm{v,st}$. 

As expected, the pure copper configuration \textit{Cu105} benefits the most from SIMD vectorization whereas the \textit{Fe126} system features the lowest vectorization speedups due to the piecewise definition of the interaction functions of the iron potential. The speedups obtained in simulations of \textit{Porous} which is also a pure copper system are somewhat lower than the speedups obtained for \textit{Cu105}. While this difference is not yet understood, we believe that it might be caused by the large number of surface atoms in this configuration which leads to a lower number of neighbors per atom. This reduces the number of force calculations per cell which in turn decreases the fraction of time spent in the vectorized parts of the code.   

The vectorization speedups obtained with the \textit{Layer} configurations are in all cases slightly lower than the corresponding speedups for \textit{Cu105} and larger than those obtained for \textit{Alloy}. This shows that our algorithm, which reloads potential parameters only when the particle type changes, improves the performance when large parts of a configuration contain only one particle type.
 
If one compares the different CPU architectures, Table~\ref{TabVec} shows that the speedups obtained with Sandy Bridge CPUs are larger than the speedups obtained with Haswell CPUs in all cases except \textit{Alloy}. This is, however, not an indication of a lower performance of the Haswell architecture. Despite the slightly lower clock frequency of the Haswell CPUs used in this work, single-threaded simulations were, compared to the Sandy Bridge processors, 15 -- 23\,\% shorter on Haswell processors without vectorization and 11 - 20\% shorter with vectorization. It is the greater increase of the speed for the non-vectorized code which causes the lower vectorization speedups on the newer CPUs. 

When comparing the behavior of the \textit{Alloy} configuration on the two Xeon architectures, it is remarkable that the vectorization speedups for this configuration are much closer to those of the \textit{Layer} system for the Haswell processors than for the Sandy Bridge processors. The reason for this difference is the expanded AVX-2 instruction set of the Haswell architecture. When we ran the executable program compiled for the Sandy Bridge machines on the Haswell processors, the (single-threaded) vectorization speedup of the \textit{Alloy} configuration dropped below the speedup of the Sandy Bridge machines. Compared to the AVX instruction set supported by the Sandy Bridge architecture, AVX-2 adds some improved SIMD support. Notably, AVX-2 includes a gather instruction which might improve the performance of the parameter reloading in simulations of multi-element systems.  

Compared to the two Xeon architectures, Table~\ref{TabVec} shows larger speedups for the Xeon Phi (Knights Corner) architecture. However, if one takes the size of the vector registers into account, the vectorization is less efficient on the Xeon Phi. This is in accordance with Amdahl's law, since the impact of the non-vectorized part of the program becomes stronger for the larger vector registers of the Xeon Phi architecture.

\begin{table*}
\caption{\label{TabProf}Comparison of CPU time spent in different sections of the force calculations. The numbers in the left column for each section are the CPU times with vectorization (v), the numbers in the right columns are times without vectorization (nv). Results were obtained from simulations using Knights Corner processors. All times are given in seconds.}
\centerline{\begin{tabular}{@{}lrrcrrcrrcrrcrrcrr@{}}\toprule
& \multicolumn{2}{c}{fetch} & & \multicolumn{2}{c}{store} & & \multicolumn{2}{c}{dist} &
& \multicolumn{2}{c}{rho} & & \multicolumn{2}{c}{embedding} & & \multicolumn{2}{c}{force}\\
\cmidrule{2-3} \cmidrule{5-6} \cmidrule{8-9} \cmidrule{11-12} \cmidrule{14-15} \cmidrule{17-18} 
& \multicolumn{1}{c}{v} & \multicolumn{1}{c}{nv} & &
\multicolumn{1}{c}{v} & \multicolumn{1}{c}{nv} & & 
\multicolumn{1}{c}{v} & \multicolumn{1}{c}{nv} & & 
\multicolumn{1}{c}{v} & \multicolumn{1}{c}{nv} & & 
\multicolumn{1}{c}{v} & \multicolumn{1}{c}{nv} & & 
\multicolumn{1}{c}{v} & \multicolumn{1}{c}{nv}\\
\midrule
\textit{Cu105}	& 43.34 & 41.33 &&	30.74 & 39.19 && 13.43 & 86.64 && 6.15 & 33.84 && 0.13 & 1.24 && 15.37 & 95.54\\
\textit{Porous}		& 15.51 & 14.42 & &	11.01 & 13.45 & &	4.46 & 30.68 & &		2.08 & 11.50 & & 	0.05 & 0.48 & &		5.20 & 31.65\\
\textit{Fe126}		& 20.62 & 20.69 & &	17.25 & 23.67 & & 	8.03 & 53.48 & &	1.62 & 7.08 & &		0.11 &  0.86 & &	54.39 & 62.77\\
\textit{Alloy}		& 65.64 & 58.68 & &	44.93 & 55.95 & &	20.16 & 127.32 & &	12.72 & 52.10 & &	0.13 & 1.27 & &		28.40 & 111.13\\
\textit{Layer}		& 49.81 & 46.34 & & 34.07 & 44.04 & &	15.09 & 98.50 & &	8.37 & 40.88 & &	0.11 & 1.08 & &		19.33 & 90.82 \\
\bottomrule
\end{tabular}}
\end{table*}
In order to better understand  the reasons for the vectorization speedups of our algorithm, we manually instrumented our program to measure the CPU time spent in different sections of the force calculations. Table~\ref{TabProf} shows the times obtained with this program on the Knight Corner machines. Section \textit{fetch} is the part of the algorithm which fetches coordinate components and calculates the components of the scaled distance vectors. Section \textit{store} stores the calculated localization functions $\rho_j$ and force components. These two sections are not vectorized. The fully vectorized section \textit{dist} transforms the scaled distance vectors into the physical coordinate system and applies boundary conditions. The sections \textit{rho} and \textit{embedding} evaluate the localization function and  the embedding energy, respectively. Finally, section \textit{force} evaluates the pair interaction and calculates the force vectors. These three sections are also vectorized.

The data in Table~\ref{TabProf} show that the non-vectorized code spends a substantial part of the simulation time in the section \textit{dist}. Vectorization reduces this time by factors between six and seven. With the Cleri and Rosatto potential \cite{Cleri:1993dn}, the sections \textit{rho}, \textit{embedding} and \textit{force} also show good vectorization speedups. For the single element systems \textit{Cu105} and \textit{Porous} these speedups are mainly limited by the evaluation of exponential functions. For the multi-element systems \textit{Alloy} and \textit{Layer}, the effectiveness of the vectorization is further reduced by the need to reload potential parameters. For \textit{Fe126} the table shows that the vectorization speedup is mainly limited by the section \textit{force}. This is understandable since this section evaluates the pair interaction for which Mendelev's iron potential~\cite{Mendelev:2003co} employs a piecewise defined function with many intervals. The localization function $\rho$ is also defined piecewise by this potential, but it uses fewer intervals and is therefore less problematic.   

From Table~\ref{TabProf} it can be seen further that with vectorization the execution time is dominated by the sections \textit{fetch} and \textit{store} for all configurations except for \textit{Fe126}. It is mainly these memory intensive sections, which access memory in an irregular pattern, that limit the SIMD vectorization speedups of simulations using the Cleri and Rosato potentials~\cite{Cleri:1993dn}.

The blocked algorithm used here for the vectorization of the force calculations and neighbor lists is different from the approach proposed by P\'{a}ll and Hess~\cite{Pall:2013gb}. Their implementation avoids the use of scatter/gather constructs altogether by grouping the atoms into small clusters. This makes it possible to load the data from all particles in a cluster into a SIMD register with a single memory transaction. The interactions between all particles in two clusters can then be evaluated with SIMD instructions. P\'{a}ll and Hess point out that for computationally light potentials such as the Lennard-Jones interaction, scatter/gather instructions should be avoided since their cost is very high. This agrees with our earlier results which showed a significant decrease in the efficiency of our vectorization method for the Lennard Jones potential~\cite{Meyer:2015ky}. For computationally more demanding potentials, our work here shows that the success of SIMD vectorization with scatter/gather instructions depends on the details of functions involved.

%
%
\subsection{Hybrid simulations on single nodes}
\label{SecSingle}
\begin{figure*}%
\centerline{\includegraphics[width=\textwidth]{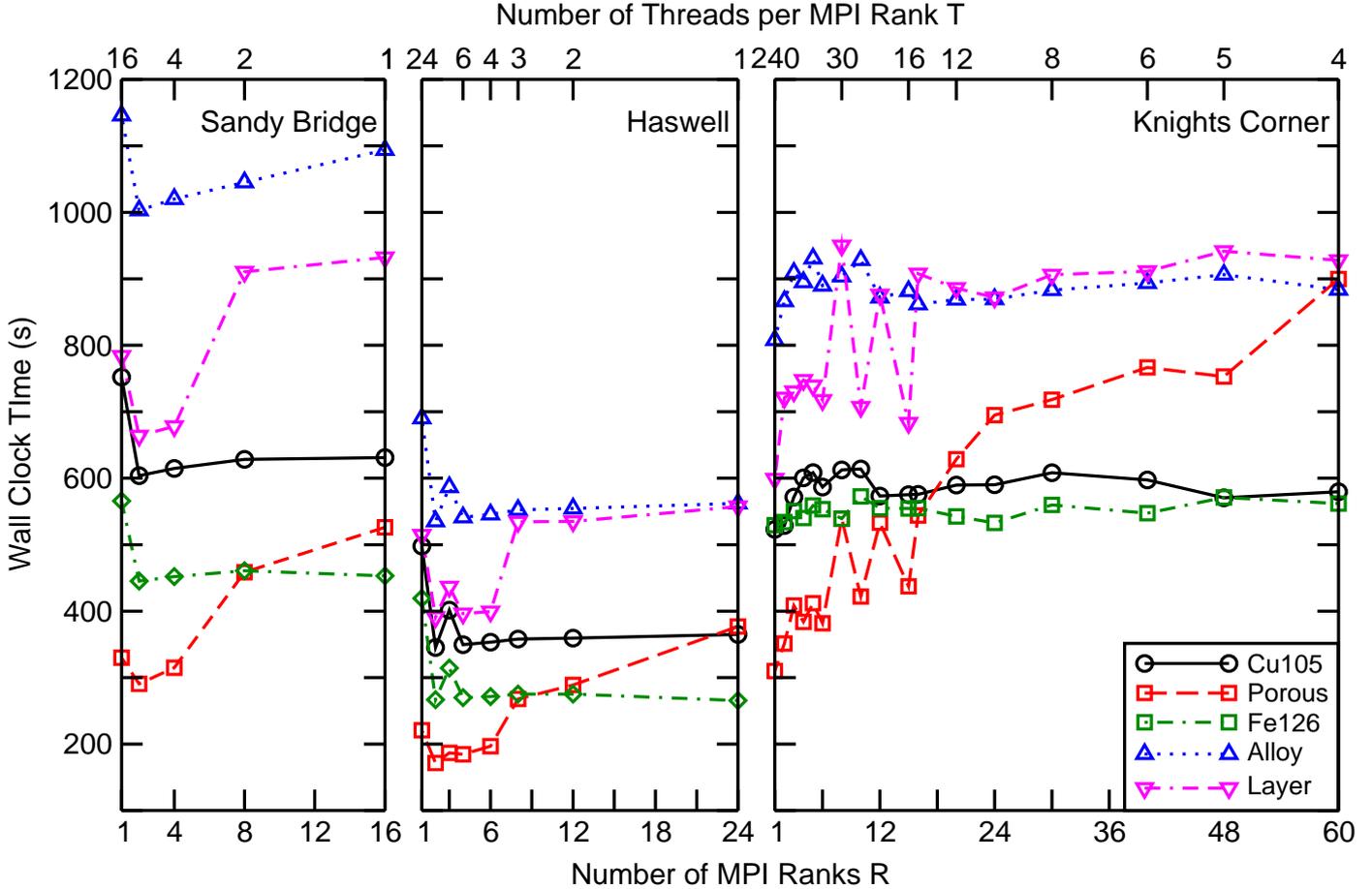}}
\caption{Simulation times for processors with Sandy Bridge, Haswell and Knights Corner architecture using the hybrid algorithm with different combinations of MPI ranks and threads. The lines between data points are only a guide to the eye.}%
\label{FigSingle}
\end{figure*}

In this section we present the results obtained with the hybrid algorithm in simulations that use all available CPU cores of a device. Figure~\ref{FigSingle} shows the average simulation time for the five benchmark configurations as a function of the number of MPI ranks $R$.  While the number of MPI ranks $R$ was varied, all simulations used the total number of cores $M$ available on the systems. To achieve this, the number of threads $T$ employed by each MPI rank was adjusted so that $R \times T = M = \mathrm{const.}$   

From Fig.~\ref{FigSingle} it can be seen that the five test configurations behave qualitatively in a similar manner on the three test architectures (notwithstanding the clear speed difference between the three systems) and that the best execution times are achieved with a low number of ranks that use a large number of threads. The relative order of the five configurations is the same on all architectures and it is the \textit{Porous} and \textit{Layer} configurations whose execution varies the most with the number of ranks $R$. This variation is a consequence of the inhomogeneity of these two configurations.  In the case of \textit{Layer} the inhomogeneity is caused by the different densities of Cu and Au. Since there are less Au atoms than Cu atoms in the same volume, the number of atoms for which a processor is responsible varies for larger numbers of MPI ranks. Due to the simple geometry of \textit{Layer}, this effect could be avoided through a careful choice of the shape of the subdomains. We chose not to do this here in order to leave this case as an example for more complicated systems where the problem cannot be avoided easily. 

For the Sandy Bridge and Haswell architectures, Fig.~\ref{FigSingle} shows a drop of the execution time when the number of ranks is increased from one to two. This is an effect of non-uniform memory access (NUMA). The computers systems used for the simulations with Xeon processors have two separate processors each of which has its own memory. The simulations shown in Fig.~\ref{FigSingle} were run with the threads of each rank pinned to a specific subset of processing cores. As far as possible these subsets were chosen such that all threads were executed on cores belonging to the same processor (close pinning). This pinning of the threads minimized the transfer of memory between the two processors. However, for one rank (and also three ranks in case of the Haswell systems), memory transfers between the processors are unavoidable. 

\begin{figure}%
\centerline{\includegraphics[width=8.5cm]{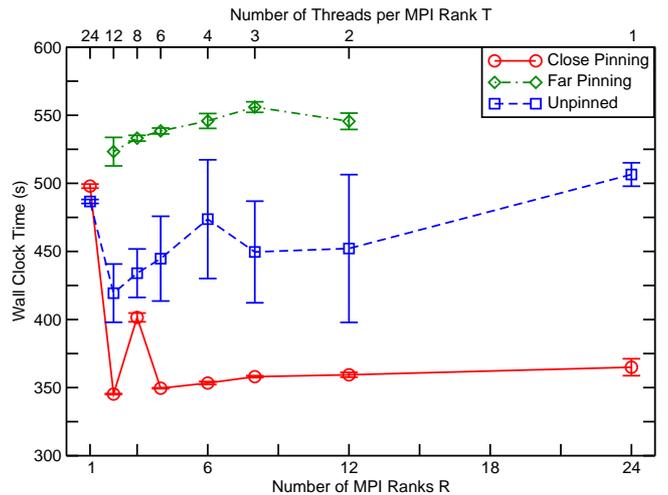}}
\caption{Simulation times of the \textit{Cu105} configuration with different pinning configurations of the threads.}%
\label{FigPin}
\end{figure}
\begin{table*}%
\caption{Maximum parallel speedups with the hybrid algorithm in simulations on a single compute node.}\label{TabHyb}
\centerline{\begin{tabular}{@{}llrrclrrclrc}
\toprule
&\multicolumn{3}{c}{Sandy Bridge} & \phantom{a} & \multicolumn{3}{c}{Haswell} & \phantom{a} & \multicolumn{3}{c}{Knights Corner}\\
\cmidrule{2-4} \cmidrule{6-8} \cmidrule{10-12}
&$S_\mathrm{max}$& $R$ & $T$& &$S_\mathrm{max}$& $R$ & $T$&  &$S_\mathrm{max}$& $R$ & $T$\\
\midrule
\textit{Cu105} 	& 25.90 & 2 & 8 & 	& 34.95 & 2 & 12 & 	& 360.06 & 1 & 240 \\
\textit{Porous}	& 22.75 & 2 & 8 & 	& 30.61 & 2 & 12 & 	& 244.42 & 1 & 240 \\
\textit{Fe126}  	& 18.11 & 2 & 8 & 	& 25.80 & 24 & 1 & 	& 212.60 & 1 & 240 \\
\textit{Alloy}     	& 22.27 & 2 & 8 & 	& 34.34 & 2 & 12 & 	& 308.44 & 1 & 240 \\
\textit{Layer}   	& 24.97 & 2 & 8 & 	& 34.00 & 2 & 12 & 	& 329.86 & 1 & 240 \\
\hline\end{tabular}}
\end{table*}
To corroborate that inter-processor memory transfers are the reason for the drop of the execution times when going from one to two ranks, we ran additional simulations where the threads were pinned in a different manner.  First, we deliberately distributed the threads of each rank over both processors (far pinning). In the case of $R=1$ or $T=1$ this type of pinning becomes identical to close pinning. For this reason we have not studied far pinning for these cases. We also ran simulations where  the threads were not pinned at all so that the operating system was free to place and move the threads (unpinned). As an example, we compare in Fig.~\ref{FigPin} the execution times of the \textit{Cu105} system for the three pinning modes. The error bars indicate the standard deviation of the execution times of five identical runs. The figure shows that the close pinning configuration results in the the lowest and most consistent execution times. With far pinning, which implies a high rate of inter-processor memory transfers, the execution times are clearly larger than in the other two cases. Running the simulations without thread pinning resulted in strongly varying, intermediate simulations times.  

On the Knights Corner systems, the localization of the memory accesses is of lesser concern. The memory system of this architecture is designed to sustain large memory bandwidths and not to depend on details of the memory placement \cite{Jeffers:13}.  As a result, memory access times are much more uniform and there is no drop of the execution times for the Knights Corner architecture when going from one to two ranks. 

The localization of the memory accesses to specific memory nodes is the reason why a hybrid algorithm can provide advantages even on a single compute node. To study this further, we give in Table~\ref{TabHyb}  the highest parallel speedups $S_\mathrm{max}$ (with respect to single-threaded simulations using the non-hybrid code without vectorization) achieved with the hybrid algorithm together with the combination of threads and ranks with which these speedups were obtained. The data show that on the Xeon systems the best speedups were obtained, in nearly all cases, in simulations with 2 ranks.  The only exception is the \textit{Fe126} configuration on the Haswell architecture. From Fig.~\ref{FigSingle} it can be seen, however, that in this case there is very little difference between the execution times for 2 or 24 ranks. Comparison of the hybrid speedups with the speedups given in Table~\ref{TabVec} reveals that in most cases the speedup provided by the hybrid algorithm outweighs the performance loss caused by the overhead of the spatial decomposition method. 

For the Knights Corner architecture, the data in Table ~\ref{TabHyb} show that the best speedups were obtained in simulations using a single rank. The hybrid algorithm provides thus  no advantage for simulations on a single Knights Corner coprocessor. The slightly higher speedup (compared to the results in Table~\ref{TabVec}) in the case of \textit{Layer} is probably caused by the different compiler version that we had to use for the Knights Corner hybrid executable.

%
%
\subsection{Multiple node hybrid simulations}
\label{SecMulti}
The principle goal of the implementation of the hybrid algorithm was to extend the applicability of the cell-task method to simulations of systems too large for a single node. This section discusses the scalability of the hybrid method in such multi-node simulations. For these experiments we used only the Haswell and Knights Corner systems.

Discussions of the scalability of parallel programs distinguish between two types of scaling: A \textit{strong scaling} test determines the time required to solve a fixed size problem as a function of the number of parallel processing units $p$. In a \textit{weak scaling} experiment, the size of the problem to be solved is not kept constant but increased by a factor of $p$ as the number of processing units is varied~\cite{HPC}. In this section we report results from both types of scaling tests.
 
In simulations running on multiple nodes the communication fabric used to exchange messages between the nodes and the associated system software are additional factors influencing execution times and speedups. In order to see the extent of this effect, we employed bulk copper configurations and phononic crystal configurations in the strong and weak scaling tests. Since in the phononic crystal configurations a considerably smaller number of atoms is close to the domain boundaries, these systems should be less affected by communication bottlenecks than the bulk systems. 

\begin{figure}%
\centerline{\includegraphics[width=8.5cm]{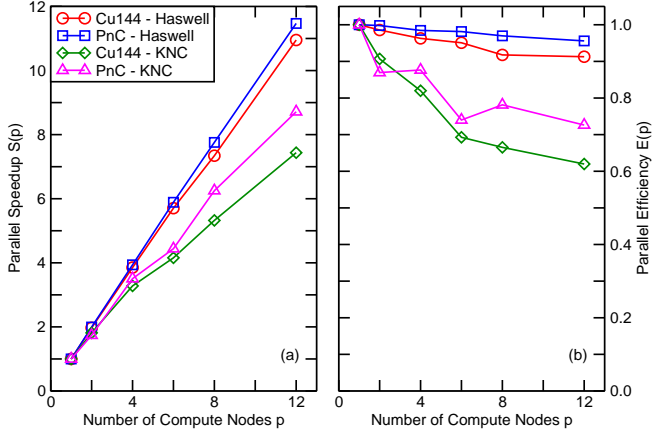}}
\caption{Strong scaling parallel speedups (left panel (a)) and parallel efficiencies (right panel (b)) in simulations of the configurations \textit{Cu144} and \textit{PnC} using multiple dual socket Haswell systems or multiple Knights Corner (KNC) coprocessors.}
\label{FigStrong}
\end{figure}
Figure~\ref{FigStrong} shows the parallel speedups and efficiencies resulting from our strong scaling experiments using up to twelve nodes with dual Haswell processors (2 x 12 cores per node) or up to twelve Knights Corner coprocessors (60 physical cores per node) to simulate the 11.9 million atom bulk system \textit{Cu144} and the 12.1 million atom phononic crystal system \textit{PnC}. The speedups were calculated with respect to simulations using a single node. The simulations on the Haswell nodes used the hybrid algorithm to run two ranks on each node with close pinning so that each 12-core processor ran one rank with 12 threads. On the Knights Corner processors, only one rank per node was used and the number of threads was varied to find the optimal number of threads. The data points shown in Fig.~\ref{FigStrong} correspond for each number of nodes to the number of threads that, averaged over five identical runs, resulted in the lowest simulation time.  

The data in Fig.~\ref{FigStrong} show a very good strong scaling behavior of the hybrid algorithm on the Haswell systems. Particularly remarkable is the behavior of the \textit{PnC} system for which the speedups remain close to the ideal $S(p)=p$ and parallel efficiencies above 95\,\% over the whole range. The efficiency for the more communication-intensive \textit{Cu144} configuration is somewhat lower but still remains above 91\,\%. 

Compared to the Haswell nodes, simulations using multiple Knights Corner systems resulted in lower parallel speedups. As shown by Fig.~\ref{FigStrong}, the parallel efficiencies drop below 80\,\% when more than 4 nodes are used. For twelve nodes the parallel efficiencies are 62\,\% and 73\,\% for the \textit{Cu144} and \textit{PnC} system, respectively. We attribute these lower parallel efficiencies to the limitations imposed by the fact that Knights Corner is a coprocessor and not a stand-alone system. 

\begin{figure}[t]%
\centerline{\includegraphics[width=8.5cm]{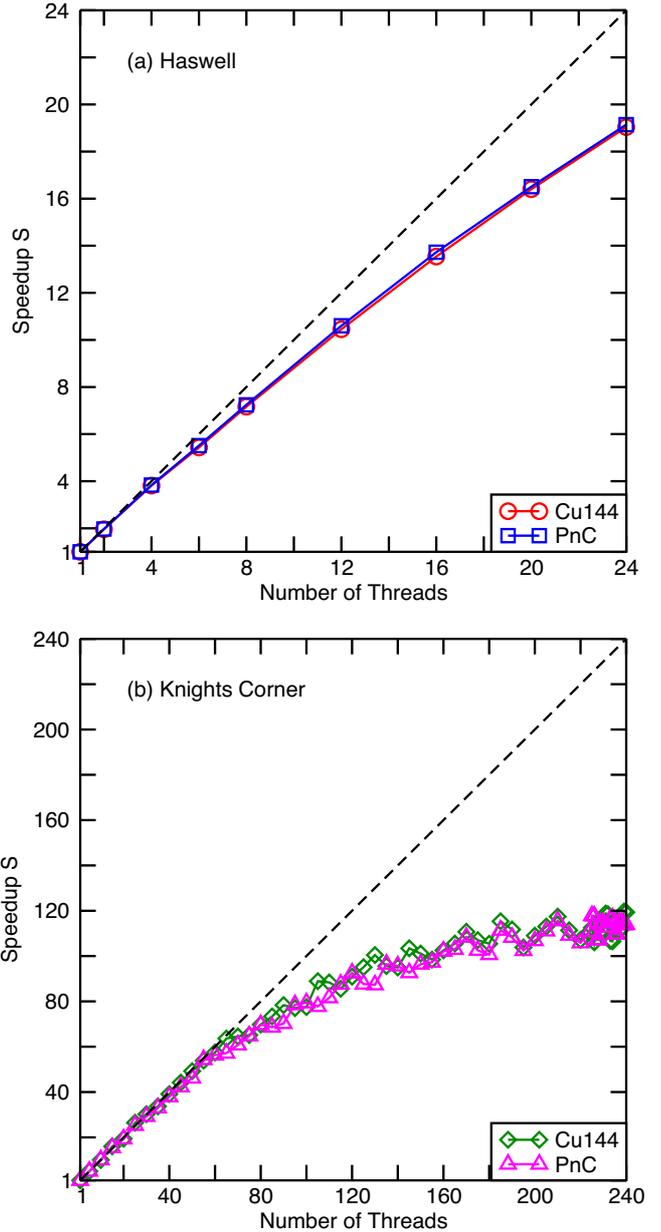}}
\caption{Strong scaling parallel speedups in simulations of the configurations \textit{Cu144} and \textit{PnC} on a dual socket Haswell system (a) and a Knights Corner coprocessor (b).}
\label{FigStrong2}
\end{figure}
In order to complement the picture of the strong scaling behavior of the hybrid algorithm, we present in Fig.~\ref{FigStrong2} single node strong scaling data. The figure shows that the two configurations obtain very similar speedups on the Haswell systems as well as on the Knights Corner processors. The maximum speedups on the Haswell nodes are 19.0 and 19.1 for the \textit{Cu144} and \textit{PnC} system, respectively. On the Knights Corner systems the maximum speedups for the two configurations are 119.6 and 117.8. The total speedups obtained for the \textit{Cu144} (\textit{PnC}) configuration with twelve compute nodes are therefore  208.5 (219.6) on the Haswell systems (288 cores in total) and  889.7 (1026.7) with the Knights Corner coprocessors (720 physical cores in total). It should be noted that on Knights Corner processors logical cores can issue new instructions only every other clock cycle~\cite{Jeffers:13}. Under ideal circumstances one therefore expects speedups of up to two times the number of physical cores.

The total speedups of 219.6 and 1026.7 achieved in the simulations of the phononic crystal configuration \textit{PnC} highlight the advantage of the hybrid version of the cell-task algorithm. On the one hand, simulations using only spatial decomposition with 288 or 1440 domains  would face severe load balancing problems due to the inhomogeneous density distribution of this configuration. The original cell-task algorithm, on the other hand, would be limited to a single compute node.   

Figure~\ref{FigStrong2} shows that on Knights Corner systems the speedups begin to oscillate when more than 60 threads are used, i.e. when there is more than one thread per core. These oscillations are probably caused by the uneven usage of resources on the cores in combination with the dynamic scheduling of the cell tasks which adds a random element. We also noted that in rare cases, the execution time drops by about 10\,\% when very few threads are used. This appears to be an effect of the placement of allocated memory on the memory banks. The effect is not present in the data shown here since the phenomenon did not occur during the final  production runs from which the data shown in Fig.~\ref{FigStrong2} were derived.

\begin{figure}%
\centerline{\includegraphics[width=8.5cm]{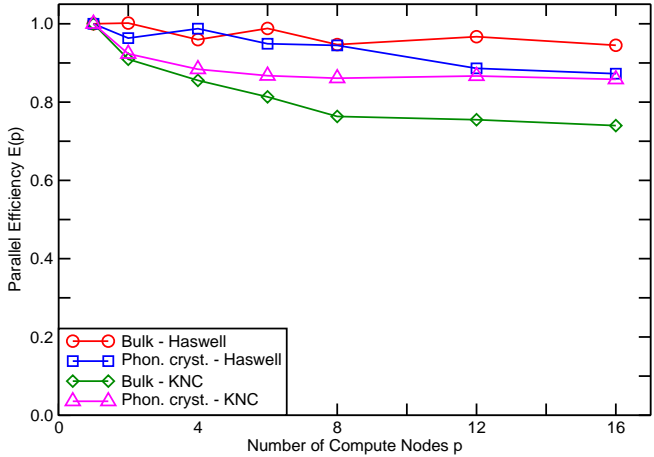}}
\caption{Weak scaling parallel parallel efficiencies in simulations of bulk and phononic crystal configurations using multiple dual socket Haswell systems or multiple Knights Corner (KNC) coprocessors.}%
\label{FigWeak}
\end{figure}

The weak-scaling parallel efficiencies of the Haswell and Knights Corner systems are shown in Fig.~\ref{FigWeak}. These efficiencies were obtained from simulations of bulk and phononic crystal configurations whose size increased with the number of compute nodes $p$.  The largest bulk and phononic crystal configurations, simulated with 16 compute nodes, contained 74.1 million and 60.7 million copper atoms, respectively.
  
As in the case of strong scaling, Knights Corner gave weak-scaling parallel efficiencies well below 100\,\%. The values are, however, significantly higher than in the strong scaling case settling quickly  around 75\,\% for \textit{Cu144} and around 85\,\% for \textit{PnC}. This settling can be explained by the fact that in the weak scaling experiments, the time for each simulation step as well as the communication volume per node remain approximately constant. As a consequence, the fraction of time spent on communication also remains constant. In contrast to this, with strong scaling, the time spent on the force calculations decreases for larger numbers of processors. This reduces the parallel efficiencies since the relative weight of the communication overhead  increases. The weak scaling parallel efficiencies of the Haswell nodes are excellent for both test configurations. This is in agreement with the good results of the strong scaling test.  

The results from multi-node simulations presented in this section show that our hybrid algorithm can be used successfully in large-scale simulations. With the Haswell nodes the parallel speedups and efficiencies are extremely good, indicating no disadvantage in the application of this method. Although the observed speedups are lower when the Knights Corner systems are used, the resulting parallel efficiencies are still acceptable.

Other hybrid parallelization schemes have shown similar results. In Ref.~\cite{Brown:2011gj}, Brown \textit{et al.} describe a hybrid parallelization method that allows the parallel use of multiple nodes with one or more GPU accelerators per node, while Pal \textit{et al.} implemented a hybrid scheme using OpenMP and MPI~\cite{Pal:2014dw}. Both articles report very high, sometimes even superlinear speedups when only normal CPUs (no GPUs) are used. This is similar to our results  with Haswell systems. The fact that we did not observe superlinear speedups is probably due to the size of our test configurations. Superlinear speedups normally occur when the problem size seen by each CPU is reduced to the point that all data fits into the CPU cache. This cannot happen with our multi-million atom configurations. In addition to this, the cell-task method was designed to work best for large inhomogeneous systems. It is not optimal for small systems. 

Using GPUs in conjunction with CPUs, the authors of Ref.~\cite{Brown:2011gj} found parallel efficiencies between 56\,\% and 102\,\% when they increased the number of  compute nodes from 1 to 15. Our parallel efficiencies for 12 Knights Corner coprocessors fall into the lower end of this range. There are, however, strong differences between our work and Ref.~\cite{Brown:2011gj} which make a direct comparison difficult. In their work, Brown \textit{et al.} used GPUs in combination with the CPUs of the host system so that only parts of the computations were performed on the GPUs. Moreover, the GPUs were addressed as accelerators using an offloading model. In contrast to this, our simulations described in this section used the Knights Corner coprocessors as stand-alone systems in native mode and did not use the host systems.

%
%
\subsection{Mixed-node simulations}\label{SecSymm}
The first generation Xeon Phis are not stand-alone processors but coprocessors which must be attached to a host system. If one uses the Xeon Phi alone, as we have done in the simulations discussed in the preceding sections, the computational power of the host system is wasted. In this section we present results from simulations that employ the hybrid algorithm to combine the computational capabilities of a host system and up to two two Xeon Phi (Knights Corner) coprocessors.

\begin{table}
\caption{\label{TabSymm}Parallel speedups in simulations of a copper bulk system ($S_\mathrm{Cu144}$) and a phononic crystal system ($S_\mathrm{PnC}$) when using the hosts processors as well as up to two Knights Corner coprocessors on the same node. $R_\mathrm{host}$, $R_\mathrm{KNC-A}$ and $R_\mathrm{KNC-B}$ are the number of MPI ranks run on the host and the two coprocessors, respectively. The speedups are relative to simulations using the host alone.} 
\centerline{\begin{tabular}{@{}lrrrrrrrrrrr}
\toprule
& \multicolumn{2}{c}{Host} & & \multicolumn{2}{c}{KNC} & & \multicolumn{3}{c}{Symmetric} \\
\cmidrule{2-3}  \cmidrule{5-6}  \cmidrule{8-10}
$S_\mathrm{Cu144}$	&  1.0 & 1.2 &	& 1.5 & 2.7 &	& 2.0   & 2.7   & 3.6 \\
$S_\mathrm{PnC}$ 		&  1.0 & 1.2 & 	& 1.5 & 2.6 &	& 1.9   & 2.7   & 3.7 \\
\midrule
$R_\mathrm{host}$		& 1     &  2	  &	& --   & -- &	& 1   & 1    & 2 \\
$R_\mathrm{KNC-A}$	& --     &  -- &	& 1   & 1  &   	& 1   & 1    & 3 \\
$R_\mathrm{KNC-B}$	& --     &  -- &	& --   & 1  &   	& --   & 1    & 3 \\
\bottomrule\end{tabular}}
\end{table}
The simulations discussed in this section are based on the same test configurations \textit{Cu144} and \textit{PnC} as the multi-node simulations in Sec.~\ref{SecMulti}. In addition to using different combinations of host processors and coprocessors, we also varied the number of MPI ranks used on each device. Table~\ref{TabSymm} gives the speedups obtained from these simulations together with the number of ranks for each device. The speedups were calculated relative to simulations that ran with a single rank on the host system alone.

Table~\ref{TabSymm} is organized in three groups. The first group (Host) shows that the usage of two ranks on the host system alone results in a speed increase of 20\,\%. This is the NUMA effect discussed in Sec.~\ref{SecSingle}. The second group (KNC) shows the speedups obtained in simulations using only the Knights Corner coprocessors. It can be seen that in our simulations a single coprocessor is 50\,\% faster than its Sandy Bridge host system. It should be noted, that this number is achieved due to the usage of SIMD vectorization. The data in Sec.~\ref{SecVec} show that we obtained vector speedups close to three on the Knights Corner architecture and close to two on processors with Sandy Bridge architecture. Without vectorization the performances would therefore be comparable. This emphasizes the need to use SIMD vectorization in order to take full advantage of the Xeon Phi architecture. With two Xeon Phi coprocessors the speedup increases to 2.7.

The third group of results in Table~\ref{TabSymm} (Symmetric) shows the results obtained when the host system and the coprocessors are used together. The table shows that the usage of the host with one or two coprocessors and one rank on each device results in speedups of 1.9 -- 2 or 2.7. These numbers are substantially lower than the combined performances of the devices. Based on the single coprocessors speedup of 1.5, the combined systems have 2.5 or 4 times the computational power of the host system alone. The lower speedups are due to the usage of a single thread per device. This leads to identical workloads on the coprocessors and the host system so that the total speed is limited by the host. 

In an attempt to balance the workload according to the capabilities of the devices we performed simulations with 2 ranks on the host system and three ranks on each of the coprocessors. This use of an unequal number of MPI ranks per node is similar to the host/device balancing by Brown \textit{et al.}~\cite{Brown:2011gj} although we did not implement dynamic load balancing. The resulting speedups of 3.6 -- 3.7 are much more satisfactory and show that the hybrid algorithm can help to overcome load balancing problems. The need to balance the workload is not limited to systems with the now outdated Knights Corner coprocessors. Although the current Knights Landing processors are only available as stand-alone processors, Intel has announced the release of Knights Landing based coprocessors at a later date. Moreover, the hybrid algorithm can serve as  a tool to balance workloads in simulations on inhomogeneous networks containing different kinds of computers.

%
%
\subsection{Future outlook: Knights Landing}
A large part of the results presented in the preceding sections were obtained on Xeon Phi coprocessors with Knights Corner architecture. When we began the work on the vectorization and hybridization algorithms, this was the only available type of Xeon Phi processors and it served us as a model for future many-core processors. With its large SIMD vector registers, hundreds of logical cores and a high bandwidth memory system the Knights Corner coprocessors allowed us to test and improve the scalability of our algorithms. 

At the time when we were completing our work, Intel released the first models of second generation Xeon Phi processors. The architecture of these processors is known as Knights Landing. This raises the question to what extent our results translate to this newer architecture. In this section, we give a partial answer to this question by presenting results that compare single node execution times and the efficiency of our vectorization algorithm on Knights Landing and Knights Corner processors. 

We did not attempt to replace all of our Knights Corner results with results obtained on Knights Landing processors for several reasons. First, the Knights Landing memory system has some NUMA features and it can be configured in different modes. In order to make claims about speedups, the effect of these modes on the performance needs to be studied.  Second, while small Knights Landing systems are becoming widely available, large scale systems that would allow us to repeat the scaling experiments of Sec.~\ref{SecMulti} are still rare and we do not have access to such a system. Since scaling data with only two or maybe four nodes are not very helpful, we have kept Knights Corner as our many-body reference platform in Sec.~\ref{SecVec} -- \ref{SecSymm}. Finally, the Knights Landing system to which we had access is shared by multiple users without a batch system. We therefore had not guarantee for exclusive access (although we did not see any hints that the data were perturbed by other users) and we could not monopolize the system. 

An important difference between all Knights Corner and the currently available Knights Landing processors is that  Knights Corner systems were only available as coprocessors whereas the current Knights Landing processors are stand-alone host processors. This difference might substantially affect the strong and weak parallel scaling of multi-node simulations. However, as already mentioned, Intel has announced the future release of Knights Landing based coprocessors so that our results from Sec.~\ref{SecMulti} and \ref{SecSymm} remain relevant.  

\begin{table*}
\caption{\label{TabKNL}Comparison of the execution times and parallel speedups of simulations on Knights Corner and Knights Landing processors. On the left, the ratio of the execution times on the two architectures is shown. $t_\mathrm{st}$, $t_\mathrm{v,st}$ ($t_\mathrm{mt}$, $t_\mathrm{v,mt}$) denote the execution times for single-threaded (multithreaded) simulations without and with vectorization, respectively. The data in the middle and on the right give the parallel speedups for both architectures similar to Table~\ref{TabVec}.}
\centerline{\begin{tabular}{@{}lllllcllllcllll@{}}\toprule
 & \multicolumn{4}{c}{$t_\mathrm{KNC}/t_\mathrm{KNL}$} & \phantom{a} &\multicolumn{4}{c}{Knights Corner} & \phantom{a}
 &\multicolumn{4}{c}{Knights Landing}\\
\cmidrule{2-5} \cmidrule{7-10} \cmidrule{12-15}
&$t_\mathrm{st}$ & $t_\mathrm{v,st}$ & $t_\mathrm{mt}$ & $t_\mathrm{v,mt}$ & 
&$S_\mathrm{v,st}$& $S_\mathrm{v,mt}$ & $S_\mathrm{mt}$ & $S_\mathrm{tot}$ &
&$S_\mathrm{v,st}$& $S_\mathrm{v,mt}$ & $S_\mathrm{mt}$ & $S_\mathrm{tot}$\\
\midrule
\textit{Cu105}	 & 3.15 & 3.88 & 2.09 & 2.13 &	 & 3.12 & 2.90 & 131.64 & 381.11 & 	 & 3.84 & 2.95 &  87.17 & 257.53 \\
\textit{Porous}	 & 3.04 & 3.38 & 2.04 & 2.13 &	 & 3.05 & 2.74 &  97.12 & 266.31 & 	 & 3.39 & 2.87 &  65.17 & 186.98 \\
\textit{Fe126}	 & 3.34 & 3.78 & 2.26 & 2.19 &	 & 1.82 & 1.73 & 129.01 & 223.47 & 	 & 2.06 & 1.68 &  87.11 & 146.43 \\
\textit{Alloy}	 & 2.79 & 3.44 & 2.06 & 2.19 &	 & 2.92 & 2.53 & 122.09 & 309.46 & 	 & 3.61 & 2.70 &  90.07 & 243.36 \\
\textit{Layer}	 & 3.02 & 3.35 & 2.07 & 2.11 &	 & 3.00 & 2.66 & 118.56 & 315.33 & 	 & 3.34 & 2.71 &  81.31 & 220.48 \\
\midrule
Ideal	 		&   &  &  &  &					& 8.00 & 8.00 &  120.00 & 960.00	&& 8.00  &  8.00  &  64.00 &  512.00 \\
\bottomrule
\end{tabular}}
\end{table*}
Table~\ref{TabKNL} compares the execution times and parallel speed\-ups obtained for the two Xeon Phi architectures. The data on the left hand side of the table show the ratio of the execution times for the five test configurations. It can be seen that in single-threaded, non-vectorized simulations the Knights Landing processors are 2.8 -- 3.3 times faster than the Knights Corner processors. With SIMD vectorization the newer architecture is 3.4 -- 3.9 times faster. The difference between vectorized and non-vectorized simulations is probably caused by an increase in the memory access speed in the newer architecture. Such an increase decreases the fraction of time spent in the non-vectorized parts of the code.  With multithreading, the speed advantage of the Knights Landing processors is reduced to factors between 2.0 and 2.3.  This can be explained by the limitations of the Knights Corner microarchitectures. As mentioned in Sec.~\ref{SecMulti}, the logical cores of Knights Corner processors can issue new instructions only every other clock cycle~\cite{Jeffers:13}. As a consequence, Knights Corner processors require at least two threads per physical core to reach peak performance and single-threaded programs run unusually slow. Since  the Knights Landing microarchitecture does not have this limitation~\cite{Jeffers:16}, multithreading speedups are lower on the newer architecture, which brings the execution times on the two architectures closer together. 

The differences between the execution times on the two architectures are reflected by the speedups shown in the middle and right part of Table~\ref{TabKNL}. It can be seen that the Knights Landing architecture gives larger vectorization speedups in single-threaded simulations whereas both architectures have similar vectorization speedups in multithreaded simulations. The Knights Corner architecture, on the other hand achieves higher speedups due to multithreading. As explained in the last paragraph, this is due to the limitations of the older architectures which limit the speed of single-threaded simulations. The speedup data for multithreaded simulations without vectorization $S_\mathrm{mt}$ show that the Knights Landing architecture achieves multithreading speedups above the number of physical cores (64) in all cases. 

%
%
\section{Summary and conclusions}
In this article, we have described a hybrid parallelization algorithm which combines two parallelization methods and an algorithm for the SIMD vectorization of molecular dynamics simulations with short-range forces. One of the main goals of the development of these algorithms was to maximize parallelism in order to keep molecular dynamics programs efficient on many-core processors with hundreds or thousands of processor cores and large SIMD vectors. An implementation of both algorithms was tested with different test configurations on standard multi-core systems as well as Knights Corner and Knights Landing  many-core processors.
 
Test simulations showed SIMD vectorization speedups of 1.3 -- 1.9 for Xeon based multi-core systems and 1.7 -- 2.9 for the Knights Corner coprocessors in multithreaded simulations. The corresponding values obtained in single-thread simulations are somewhat higher. These speedups are significantly lower than the sizes of the SIMD registers which is four for the multi-core systems and eight for Knights Corner. An analysis of the speedups of individual parts of the force calculation, showed that the effect of SIMD vectorization is limited by the code sections that fetch and store data from and to memory. Due to the irregular memory access pattern of molecular dynamics simulations, these parts of the code cannot be vectorized efficiently.    
The lowest vectorization speedups were obtained in the simulations of bulk iron using the potential by Mendelev~\cite{Mendelev:2003co}. This is understandable since this potential uses piecewise defined functions that are hard to vectorize. It can be expected that simulations using a tabulated potential will face similar problems since the use of an interpolation table is equivalent to a piecewise defined function. Although the vectorization speedups remain significantly behind their theoretical limits, the obtained speedups still represent substantial reductions in simulation time. 

Simulations with the hybrid algorithm on a single compute node showed that on dual socket Xeon systems, the usage of two ranks whose threads are pinned to separate CPU sockets performs significantly better than spatial decomposition or the cell-task method alone. The hybrid algorithm was not beneficial in simulations on a single Knights Corner coprocessor. These results show that the hybrid algorithm can be used to optimize the simulation speed on NUMA systems through the localization of each MPI rank's memory accesses to the closest memory node.

Strong and weak scaling tests revealed extremely good scaling of the hybrid algorithm on the Xeon based multi-core systems. With the Knights Corner systems we found a less impressive but still acceptable scaling behavior. We attribute this to the limitations of the inter-node communication caused by the coprocessor nature of the Knights Corner systems. The good parallel efficiencies obtained in simulations of phononic crystal systems with sizes up to 60 million atoms highlight the potential of the hybrid extension of the cell-task method to enable efficient large-scale simulations of complex inhomogeneous systems. 

Simulations performed on the combination of a Sandy Bridge host system with one or two Knights Corner coprocessors resulted in speedups of up to 3.7 compared to simulations  on the host system alone. This is an example for the usefulness of the hybrid algorithm as a tool to perform parallel simulations on different devices with unequal computing powers.   

Preliminary results obtained on Knights Landing processors show that the SIMD vectorization results obtained with Knights Corner processors carry over to the newer generation of Xeon Phi processors. While we could not yet perform such experiments, we expect that strong and weak scaling tests on systems with stand-alone Knights Landing processors will result in higher parallel efficiencies than those obtained with Knights Corner coprocessors. The memory architectures and on-die interconnect systems of Knights Landing and Knights Corner processors differ substantially. The behavior of these systems on Knights Landing processors can be controlled through two configuration parameters ``Cluster Mode" and ``Memory Mode". We expect that, depending on the settings of these parameters, usage of the hybrid algorithm with multiple ranks per processors will be beneficial on Knights Landing processors. 

The full computational power of many-core processors such as the Xeon Phis can only be realized with programs that support high levels of parallelism as well as SIMD vectorization. Our work shows how this can be achieved in short-range molecular dynamics simulation programs. This makes current and future many-core processors attractive systems for large-scale simulation studies using 50 million atoms or more. 

\section*{Acknowledgments}
This work has been made possible by the Compute/ Calcul Canada facilities at Calcul Qu\'{e}bec, SciNET and SHARCNET. Financial support by Laurentian University and the  Natural Science and Engineering Research Council of Canada (NSERC) [grant number 371446-11]  is gratefully acknowledged. C.M. has been supported through the Ontario Graduate Scholarship (OGS) program. 
\section*{References}

\end{document}